\documentclass[prb,aps,a4paper,twocolumn,showpacs,superscriptaddress,floatfix,10pt]{revtex4-1}
\usepackage{graphicx,amsfonts,amssymb,amsmath,xspace,dsfont}
\usepackage[colorlinks=true,citecolor=green,linkcolor=red,urlcolor=blue]{hyperref}
\usepackage{color}

\definecolor{marc}{rgb}{.27,.51,.70}

\newcommand{\kagome}{Kagom\'e}
\newcommand{\picket}[1]{\left|\begin{array}{c}\!\!\includegraphics[width=.5cm]{#1}\!\!\end{array}\right\rangle }
\newcommand{\picbra}[1]{\left\langle\begin{array}{c}\!\!\includegraphics[width=.5cm]{#1}\!\!\end{array}\right| }

\begin{document}

\allowdisplaybreaks
\title{Extend of the \texorpdfstring{$\mathbb{Z}_2$}{Z(2)} spin liquid phase on the \kagome-lattice}
\author{Marc D. Schulz}
\email{marcdaniel.schulz@tu-dortmund.de}
\affiliation{TU Dortmund}

\begin{abstract}
The $\mathbb{Z}_2$ topological phase in the quantum dimer model on the \kagome -lattice is a candidate for the description of the low-energy physics of the anti-ferromagnetic Heisenberg model on the same lattice. We study the extend of the topological phase by interpolating between the exactly solvable parent Hamiltonian of the topological phase and an effective low-energy description of the Heisenberg model in terms of a quantum-dimer Hamiltonian. Therefore, we perform a perturbative treatment of the low-energy excitations in the topological phase including free and interacting quasi-particles. We find a phase transition out of the topological phase far from the Heisenberg point. The resulting phase is characterized by a spontaneously broken rotational symmetry and a unit cell involving six sites.
\end{abstract}

\maketitle

\section{Introduction}
In the last decades, there has been an increasing interested in topologically ordered phases of matter. 
One manifestation of this exotic order in magnetic systems are the so called spin-liquids. 

The most promising candidate for the realization of this type of phase in nature is the compound \textit{Herbertsmithite} ($\rm ZnCu_3(OH)_6Cl_2$), for which the relevant magnetic moments are arranged in two-dimensional layers forming a \kagome -lattice\cite{shores05,han12,norman16}.

However, while experimental findings provide increasing evidence for the presence of the spin-liquid phase\cite{imai16,sherman16}, theoretical predictions for the relevant anti-ferromagnetic spin-$\frac{1}{2}$ Heisenberg model remain inconclusive as there are various competing states present in the low-energy regime of this model. The candidate phases of matter include the already mentioned spin liquids (gapped\cite{yan11,depenbrock12,jiang12,mei17} and gapless\cite{ran07,iqbal13,iqbal14,he17,liao17}) as well more conventionally phase such as Valence-Bond Solids (VBS) with unit cells of six\cite{ran04}, twelve\cite{syromyatnikov02}, eighteen\cite{marston91} or thirty-six sites\cite{marston91,nikolic03,singh07}. 

In order to enable large-scale numerical studies for this frustrated model, effective low-energy models in terms of quantum dimers have been derived in the past\cite{zeng95,misguich03,poilblanc10,rousochatzakis14}. Despite their reduced complexity, the investigation of these models still proves to be rather challenging.\cite{poilblanc11}

Interestingly, also the quantum dimer model on the \kagome -lattice hosts a spin-liquid phase in a certain parameter regime.\cite{misguich02} Previous studies did investigate the extend of this phase as well as phases arising beyond the phase transition.\cite{schwandt10,poilblanc11,hao14,hwang15,ralko17} In the case of Refs.~\onlinecite{schwandt10,poilblanc11}, this analysis was performed by means of exact-diagonalization techniques. Their results indicate a transition to a thirty-six site VBS-phase in the close vicinity to the point corresponding to the Heisenberg anti-ferromagnet. \\
However, this method introduces additional approximations, notably finite-size effects. To provide a complementary approach, we perform in this work an analysis based on techniques yielding directly results in the thermodynamic limit and compare those with the previously reported ones.

Therefore this work is organized as follows: We present the quantum-dimer model with the relevant parameters in Sec.~\ref{sec:model}. Our approach to yield an effective low-energy description is detailed in Sec.~\ref{sec:pcut}. In Sec.~\ref{sec:results}, we present our results for the extend of the topological phase as well as estimates for the non-topological phase beyond the transition. Our discussion of the results in Sec.~\ref{sec:conclusion} concludes this work.

\section{Model}\label{sec:model}
The model of (hardcore) dimers on the \kagome-lattice arises as the low-energy description of the anti-ferromagnetic Heisenberg model on the same lattice. The hardcore dimers result from a representation of singlets formed by two spin-$\frac{1}{2}$ of the original anti-ferromagnet. In Sec.~\ref{ssec:qdm}, we introduce a general quantum-dimer Hamiltonian $H_{\rm qdm}$ and specify its parameters to yield the Hamiltonian $H_{\rm QDM}$ for the low-energy description of the Heisenberg model\cite{poilblanc10,schwandt10}.
 In Sec.~\ref{ssec:z2ham}, we discuss the model $H_{\rm qdm}$ for a different set of parameters. For those, the resulting model $H_{\mathbb{Z}_2}$ is exactly solvable and realizes a topologically ordered phase. We use this exactly solvable Hamiltonian as a starting point and study its properties as we interpolate between the starting point and the quantum-dimer Hamiltonian $H_{\rm QDM}$. We discuss the details of our analysis in Sec.~\ref{ssec:interpol}. 

\subsection{The quantum dimer model}\label{ssec:qdm}
The Hilbert space of this effective model is given by dimer coverings of the \kagome -lattice, i.e. two nearest-neighbor sites are forming a dimer as depicted, e.g., in Fig.~\ref{fig:refstate}. Within this Hilbert space, the general quantum-dimer Hamiltonian $H_{\rm qdm}$ consists of two families of terms: The first one (kinetic or resonance terms) changes the position of dimers along closed loops connecting the sites of the lattice. The second family (potential terms) measures whether such a change is possible. 
In this work, we consider the quantum-dimer model derived in Ref.~\onlinecite{schwandt10}, which contains kinetic as well as potential terms involving loops around a single hexagonal plaquette of the \kagome -lattice. The corresponding Hamiltonian $H_{\rm qdm}$ reads
\begin{align}
H_{\rm qdm}=&J_6 \picket{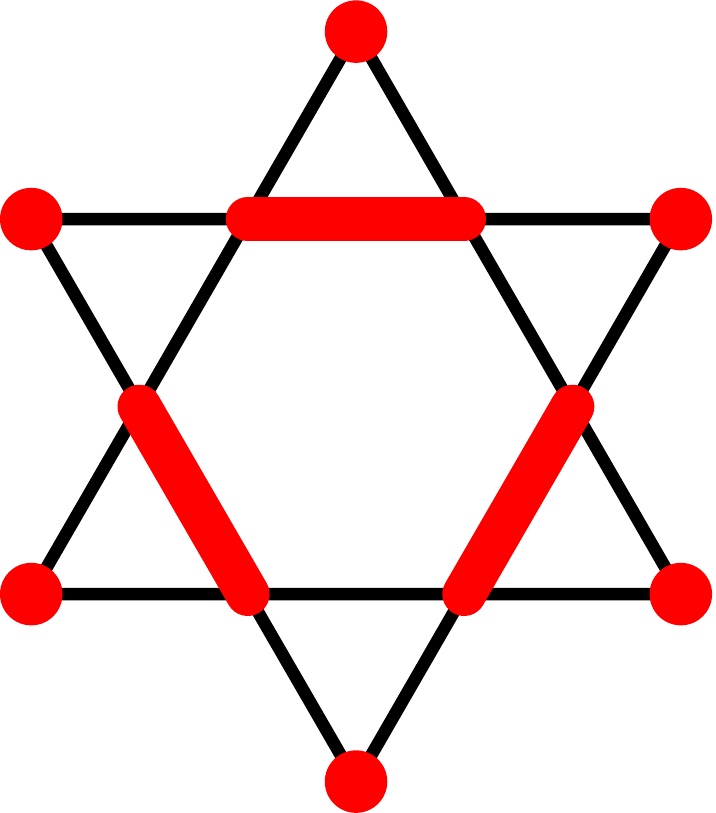}\!\!\picbra{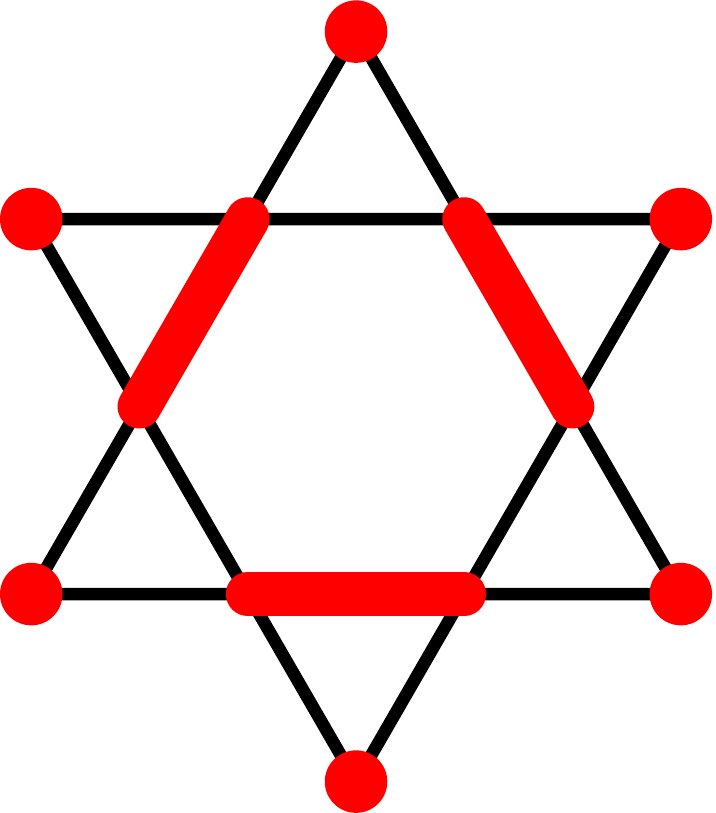}+V_6 \picket{./dimer_D_1b.pdf}\!\!\picbra{./dimer_D_1b.pdf}\nonumber\\+
&J_8 \left( \picket{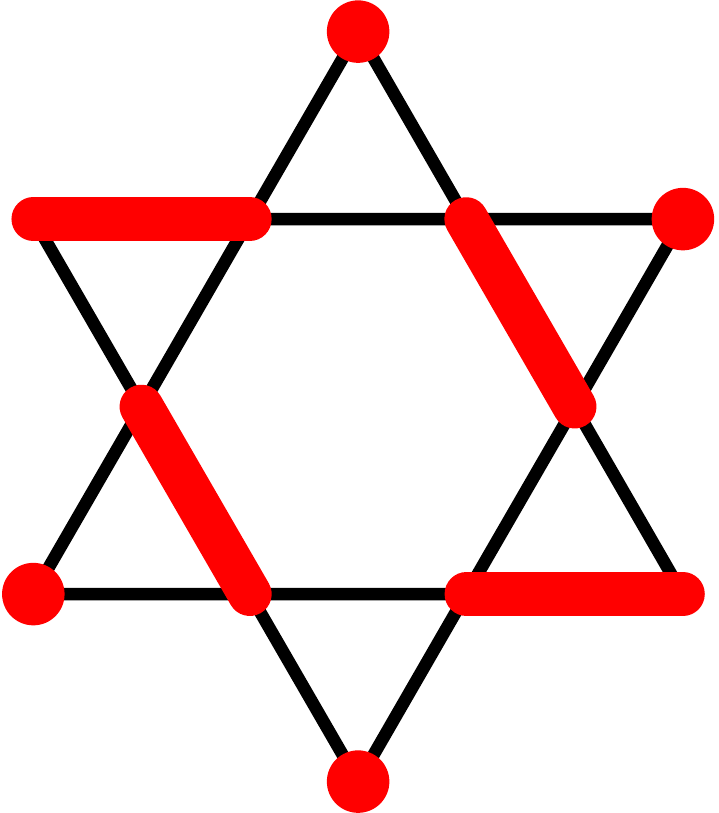}\!\!\picbra{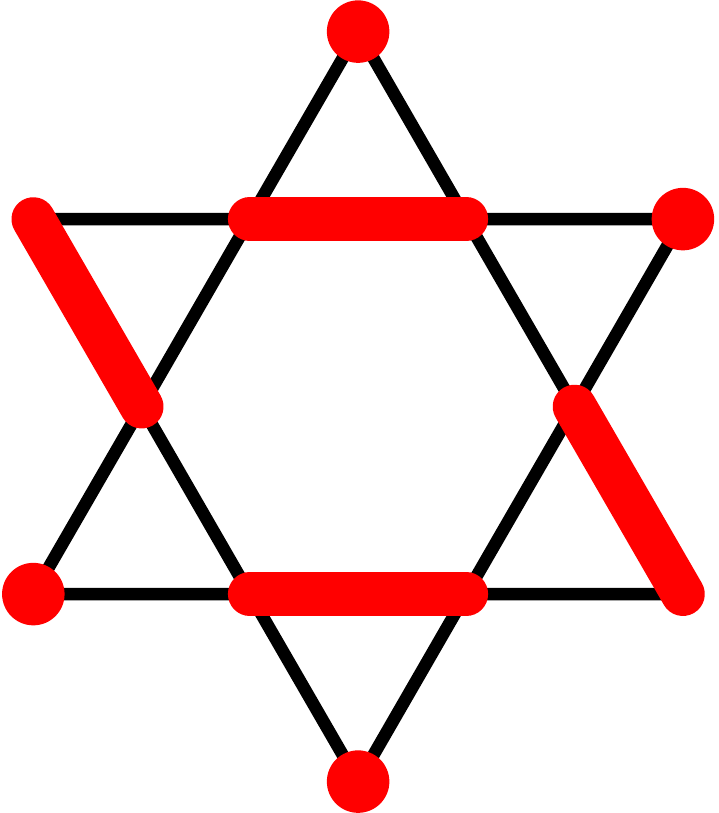}+\picket{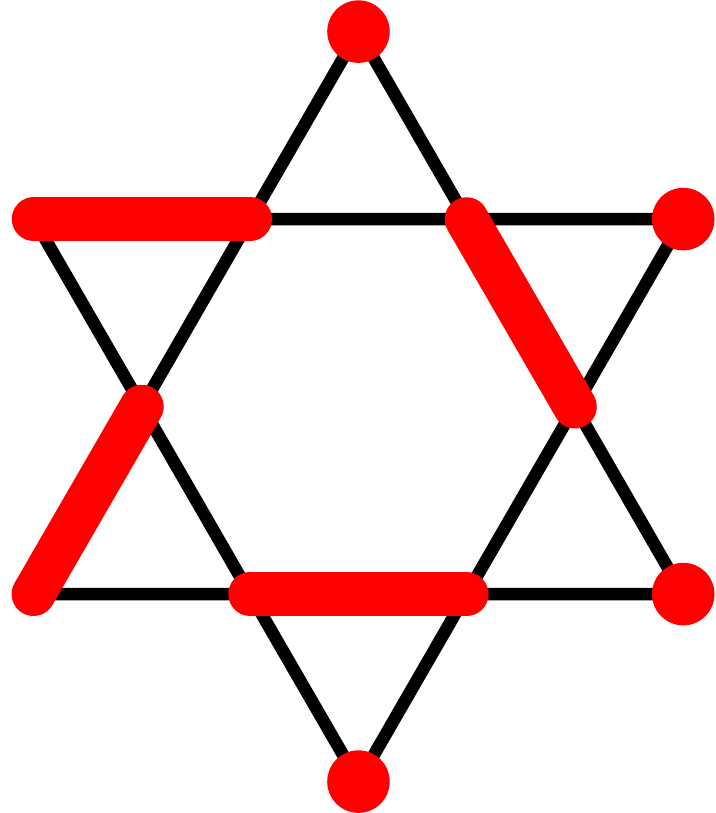}\!\!\picbra{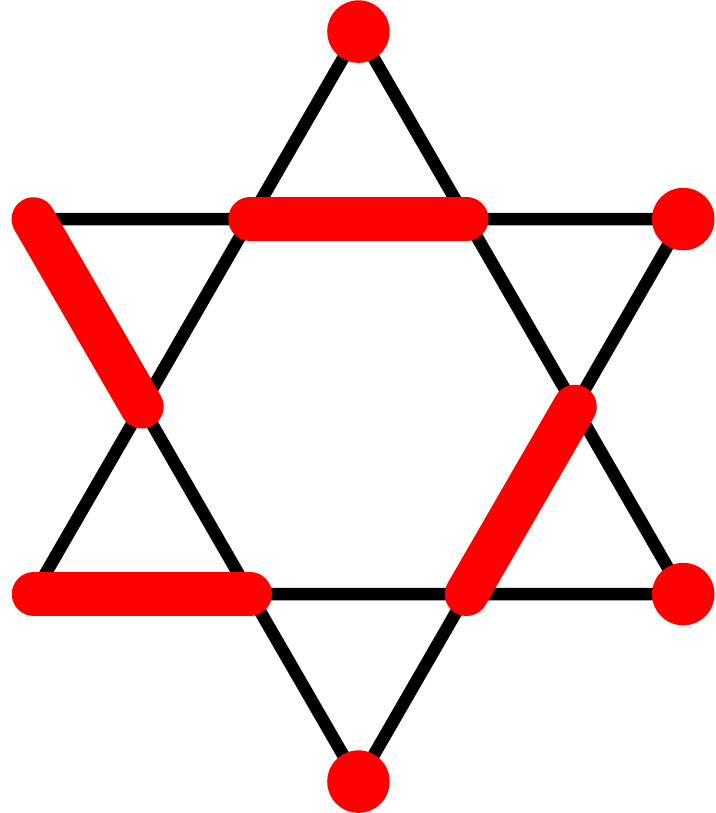}+\picket{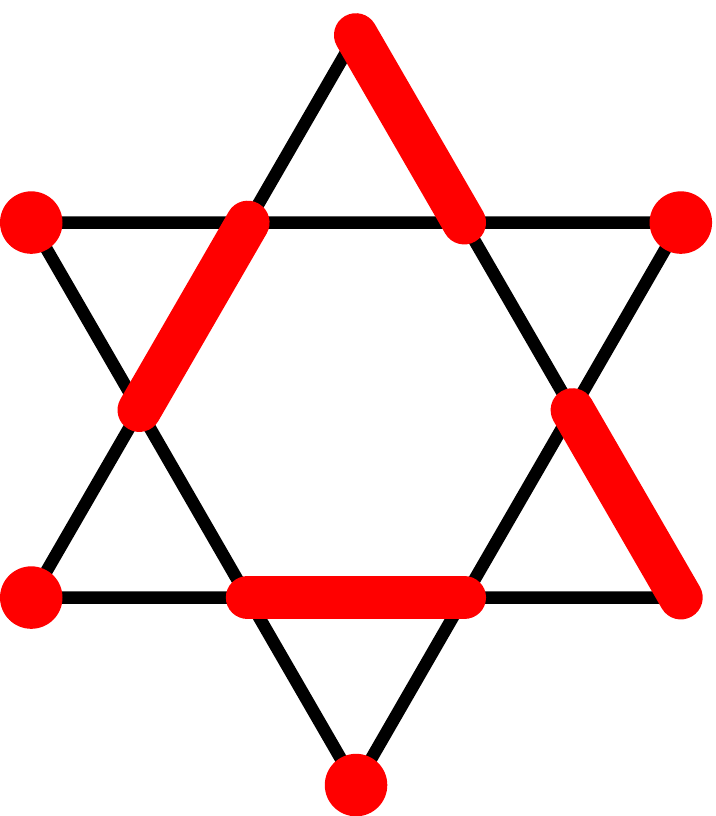}\!\!\picbra{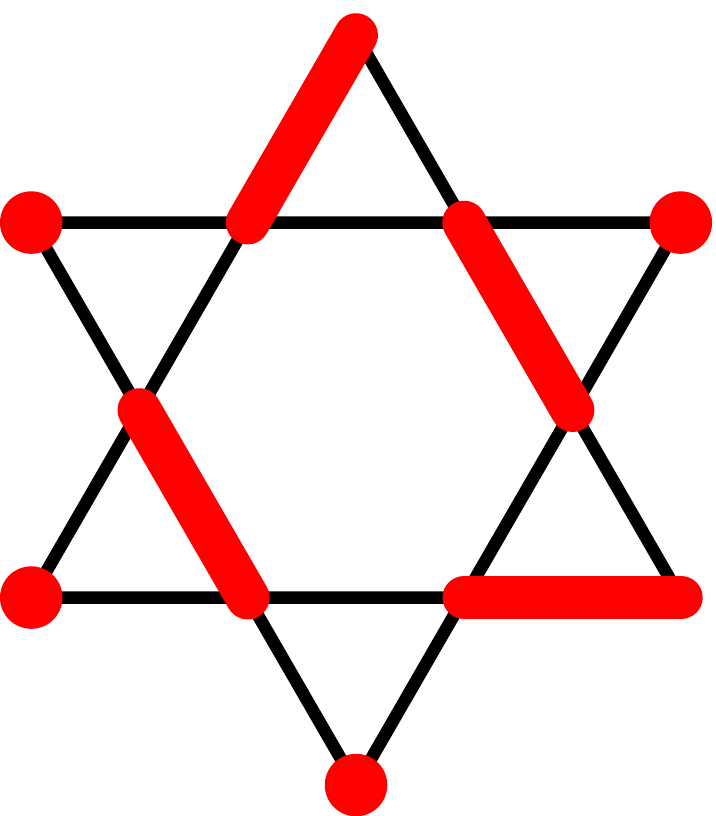}\right)\nonumber\\+
&V_8 \left( \picket{./dimer_C1_1b.pdf}\!\!\picbra{./dimer_C1_1b.pdf}+\picket{./dimer_C2_1b.pdf}\!\!\picbra{./dimer_C2_1b.pdf}+\picket{./dimer_C3_1b.pdf}\!\!\picbra{./dimer_C3_1b.pdf}\right)\nonumber\\
+&J_{10}\! \left( \picket{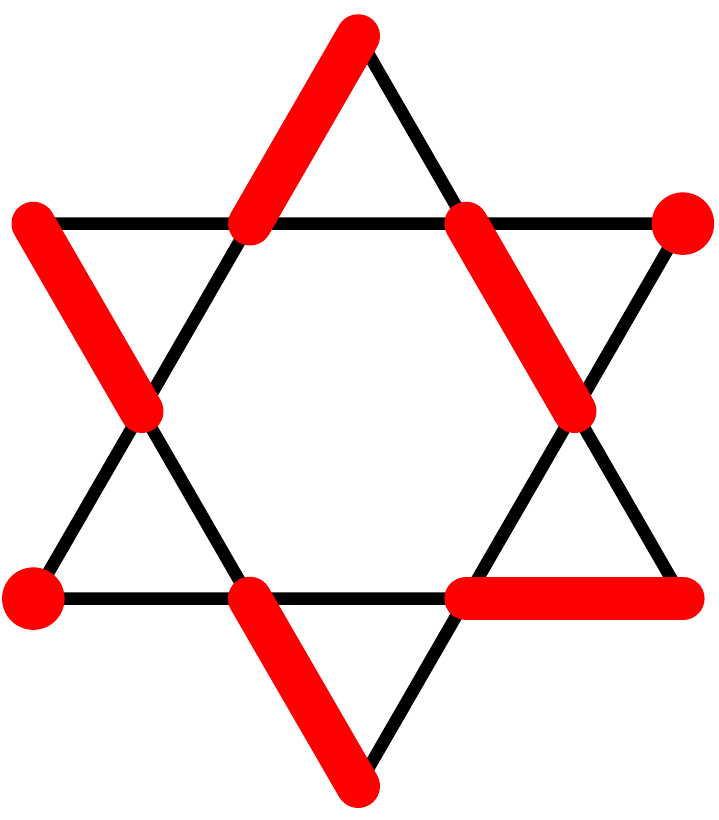}\!\!\picbra{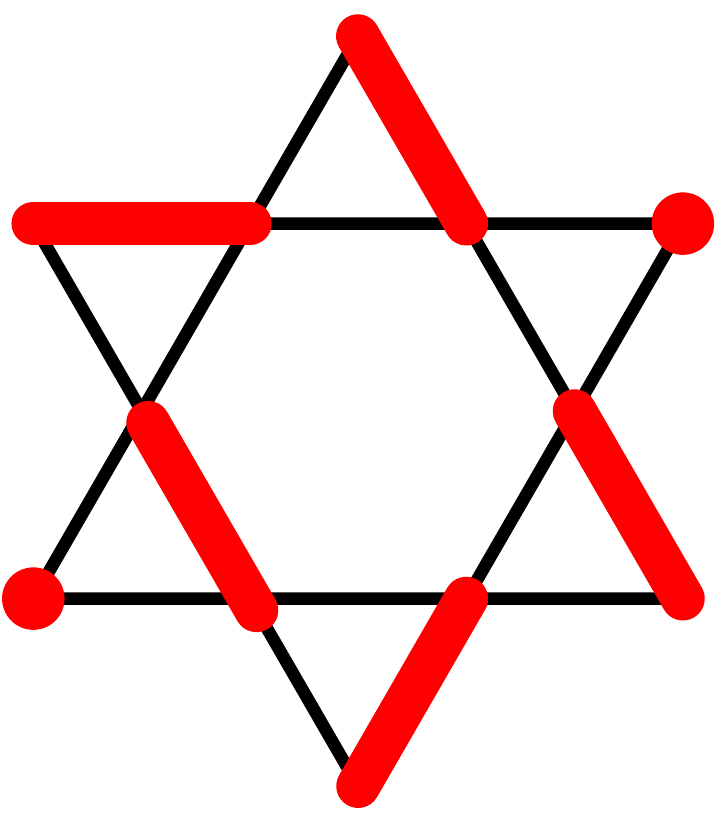}+\picket{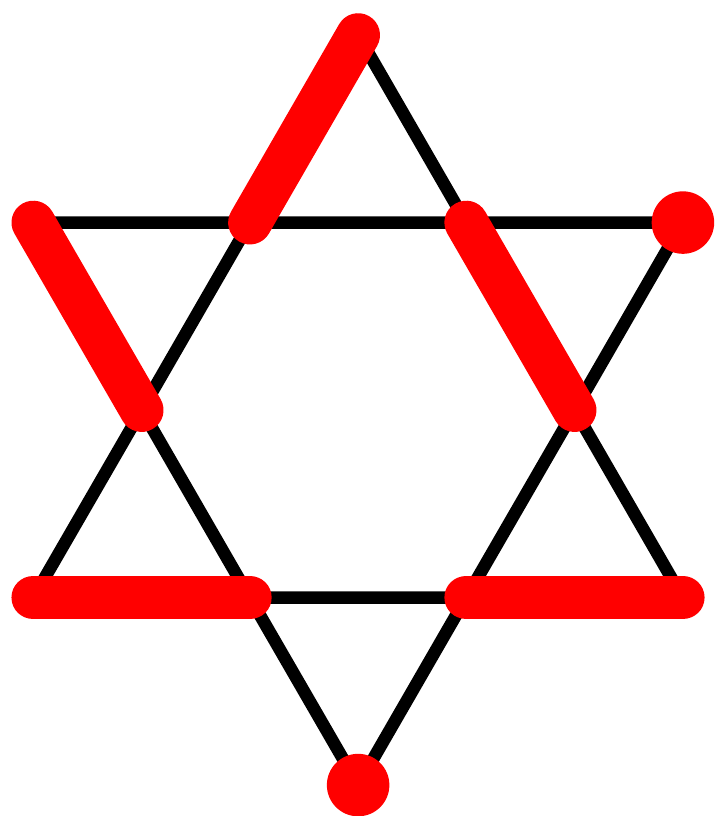}\!\!\picbra{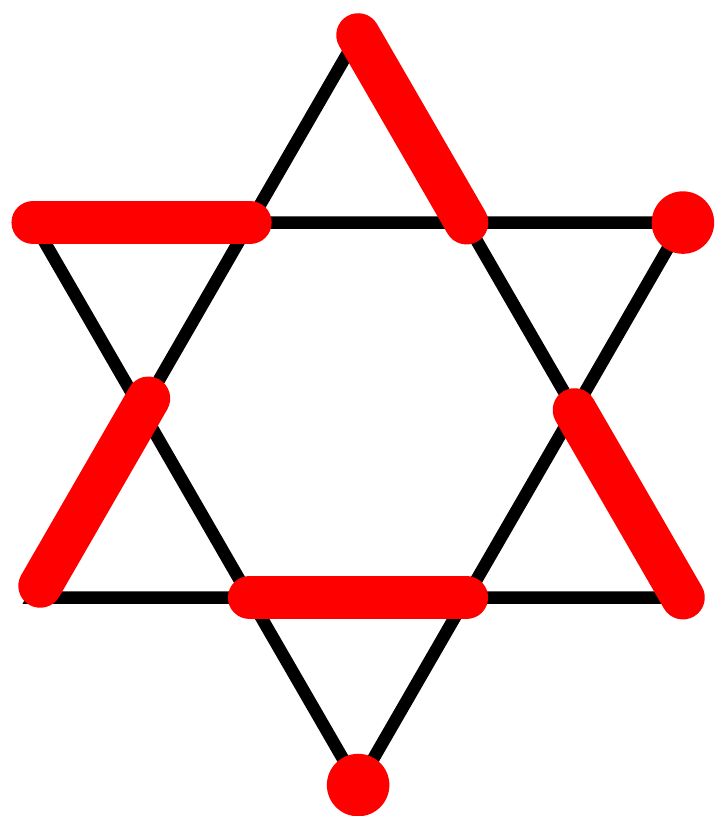}+\picket{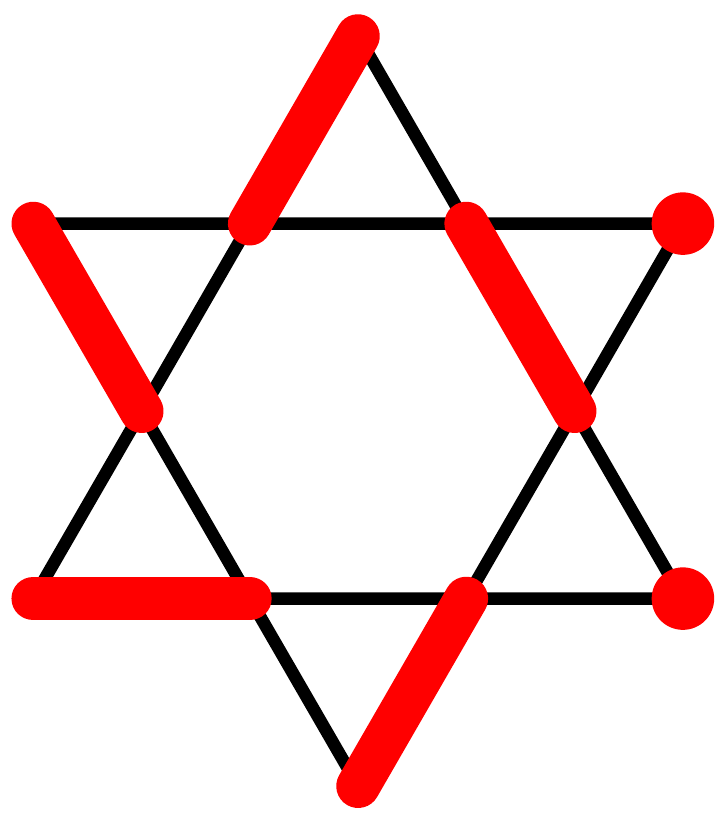}\!\!\picbra{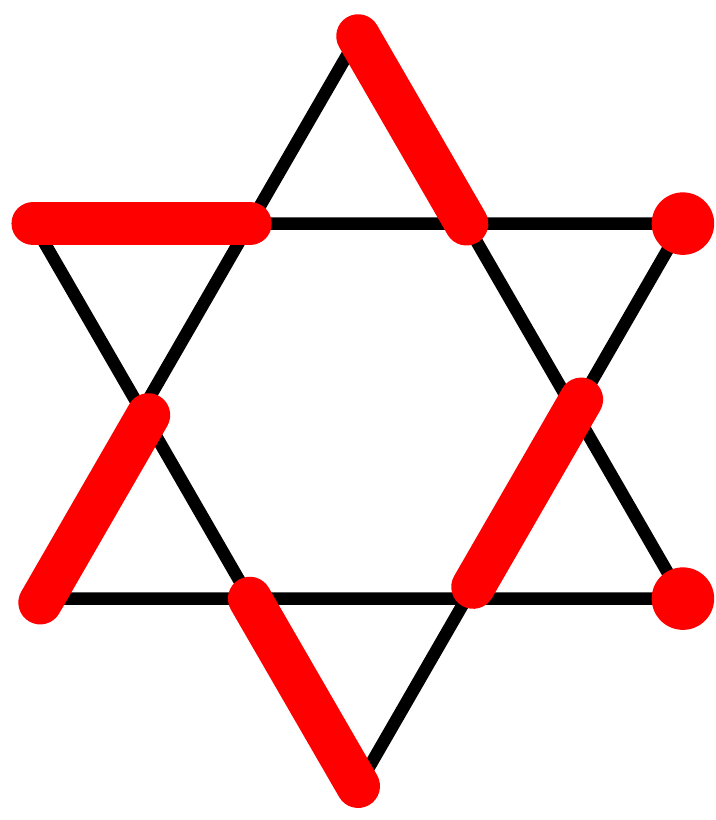}\right)\!\nonumber\\
+&V_{10}\!\left(\picket{./dimer_B1_1b.pdf}\!\!\picbra{./dimer_B1_1b.pdf}+\picket{./dimer_B2_1b.pdf}\!\!\picbra{./dimer_B2_1b.pdf}+\picket{./dimer_B3_1b.pdf}\!\!\picbra{./dimer_B3_1b.pdf}\right) \!\nonumber\\
+&J_{12} \picket{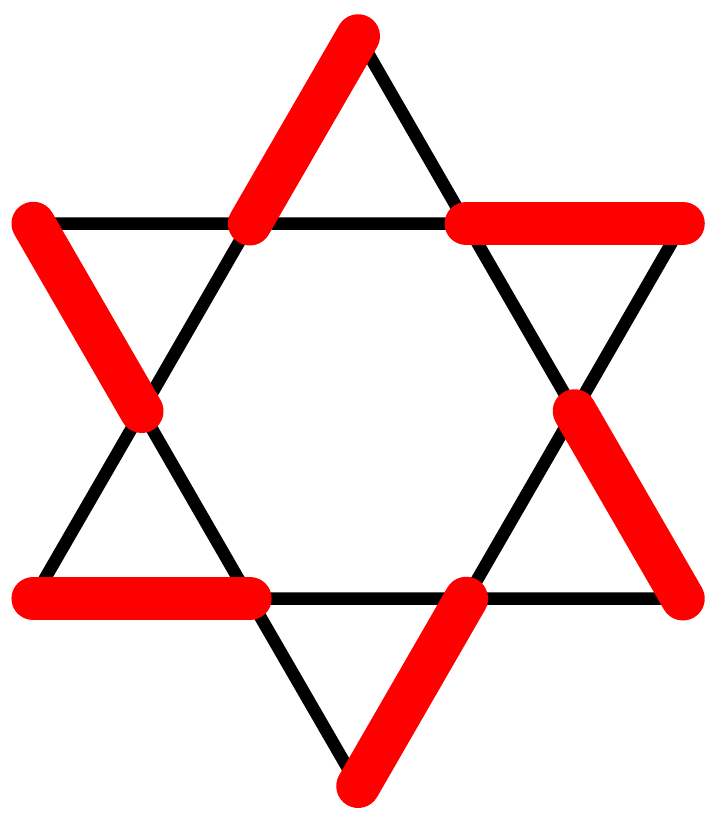}\!\!\picbra{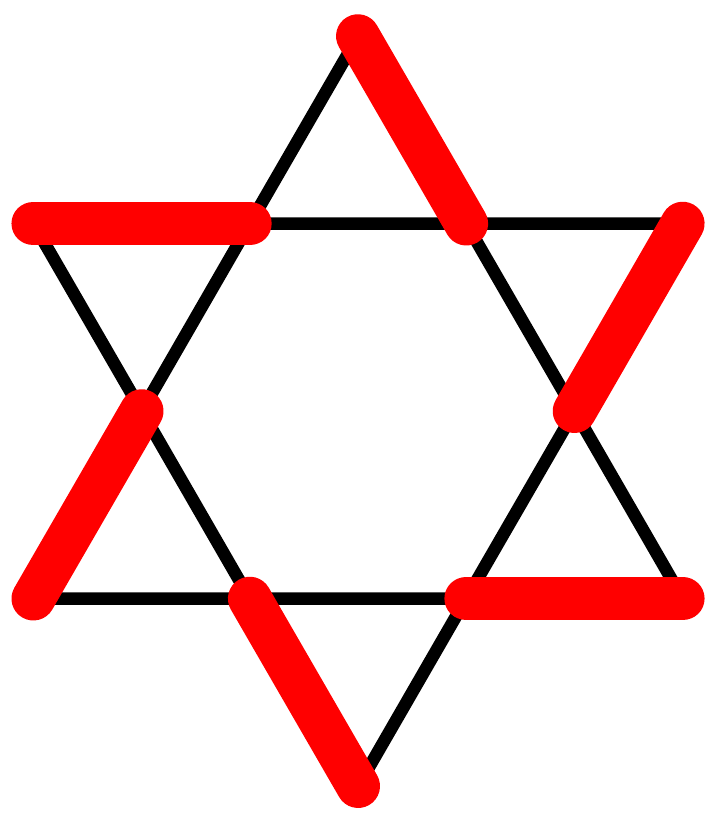}+V_{12} \picket{./dimer_A_1b.pdf}\!\!\picbra{./dimer_A_1b.pdf},
\label{eq:ham_qdm}
\end{align}
where each term is representing also rotated and translated versions of the respective dimer-configuration as well as versions which can be connected to the previous by flipping the dimers along a closed loop around the depicted plaquette. The value of the couplings given in Ref.~\onlinecite{schwandt10} are
\begin{align}
\quad\quad & J_{6}=-\frac{4}{5}, & J_{8}=\frac{16}{63},\ \ & J_{10}=-\frac{16}{255}, & J_{12}=0, \nonumber\\
\quad\quad & V_{6}=\phantom{-}\frac{1}{5}, & V_{8}=\frac{2}{63},\ \ & V_{10}=\phantom{-}\frac{1}{255}, & V_{12}=0. 
\label{eq:smp_couplings}
\end{align}
We will label the quantum-dimer Hamiltonian $H_{\rm qdm}$ (\ref{eq:ham_qdm}) for these values as $H_{\rm QDM}$ in the following. The goal of this work is to shed light on the phase realized in the low-energy regime of $H_{\rm QDM}$.

\subsection{The \texorpdfstring{$\mathbb{Z}_2$}{Z(2)}-Hamiltonian}\label{ssec:z2ham}
However, the starting point of our analysis is another point in the parameter space of the quantum dimer model (\ref{eq:ham_qdm}). For the parameter values {$J_6=J_8=J_{10}=J_{12}=-\frac{1}{4}$} and $V_6=V_8=V_{10}=V_{12}=0$, the quantum-dimer Hamiltonian $H_{\rm qdm}$ takes the form
\begin{align}
H_{\mathbb{Z}_2}=-\frac{1}{4} \sum_p B_p,
\label{eq:ham_z2}
\end{align}
where the operator $B_p$ sums up all the kinetic terms involving the plaquette $p$ in Eq.~(\ref{eq:ham_qdm}). The Hamiltonian (\ref{eq:ham_z2}) has been introduced, e.g., in Ref.~\onlinecite{misguich02}.

The Hamiltonian $H_{\mathbb{Z}_2}$ is exactly solvable. This can be inferred from the following properties of the operators $B_p$:
We have $\left[B_{p^{\phantom{\prime}}},B_{p^{\prime}} \right]=0$, i.e.~operators on different plaquettes $p^{\phantom{\prime}}\!$, $p^{\prime}$ commute. The spectrum of the operator $B_p$ consists of the eigenvalues $\pm 1$, which can be seen, e.g., from the relation $B_p^2=\mathds{1}$.
Using these two properties, we can construct the operators 
\begin{align}
P^{(0)}_p&=\frac{1}{2}\left(\mathds{1}+B_p\right) \text{ and }\label{eq:nofluxproj}\\
P^{(1)}_p&=\frac{1}{2}\left(\mathds{1}-B_p\right). \label{eq:fluxproj}
\end{align}
The operator $P^{(0)}_p$ ($P^{(1)}_p$) projects onto eigenstates of $B_p$ with eigenvalue $+1$ ($-1$). As the $B_p$, the projectors $P^{(0)}_{p^{\phantom{\prime}}}$ and $P^{(1)}_{p^{\prime}}$ commute for different $p^{\phantom{\prime}}\!$, $p^{\prime}$.

\begin{figure}[ht]%
\includegraphics[width=.45\columnwidth]{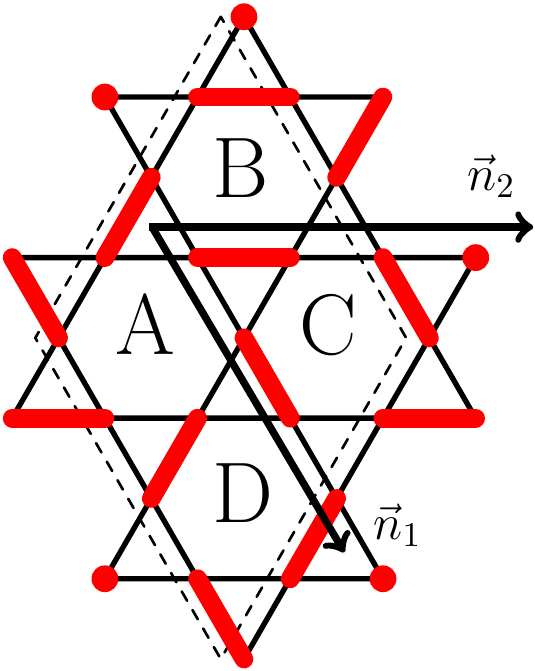}%
\caption{The reference state $\left| {\rm ref}\right\rangle$ is depicted in the four-plaquette (or twelve-site) unit cell spanned by the basis vectors $\vec{n}_1$ and $\vec{n}_2$. The labels $\rm A$,$\rm B$,$\rm C$, and $\rm D$ denote the different hexagonal plaquettes within this unit cell. The single dots denote here end-points of dimers not fully fitting on the depicted lattice.} %
\label{fig:refstate}%
\end{figure}

We can thus see that the Hamiltonian (\ref{eq:ham_z2}) is essentially the sum of commuting projectors $P^{(1)}_p$.
Eigenstates of the operators $P^{(0)}_p$ ($P^{(1)}_p$) with eigenvalue $+1$ are said to host no flux (a flux) on plaquette $p$.
Using those projectors, we can construct eigenstates of the Hamiltonian (\ref{eq:ham_z2}) similarly as detailed, e.g., in Refs.~\onlinecite{vidal08,schulz12}:
A state with fluxes on the plaquettes ${\bf P}=\lbrace p_1,p_2,\ldots,p_n\rbrace$ can be represented as 
\begin{align}
\left|\lbrace p_1,p_2,\ldots,p_n\rbrace \right\rangle = \mathcal{N} \prod_{p\in {\bf P} } P^{(1)}_p\prod_{p\notin {\bf P} } P^{(0)}_p \left| {\rm ref}\right\rangle,
\label{eq:example_state}
\end{align}
where $\mathcal{N}$ is a normalization factor and we choose the reference state $\left| {\rm ref}\right\rangle$ as depicted in Fig.~\ref{fig:refstate}.

The ground state of the Hamiltonian $H_{\mathbb{Z}_2}$ is flux-free. One can work out that it is the equal-amplitude superposition of all possible dimer configurations. The presence of a flux on plaquette $p$ corresponds to a phase of $\pi$ between configurations only differing by the action of a kinetic term on plaquette $p$.

If defined on an open plane, the flux-eigenstates (\ref{eq:example_state}) form a complete orthogonal basis-set. If defined on a non-trivial topology, additional non-local quantum numbers are needed to distinguish different states. As we work on the trivial topology of the open plane in this work, we refer, e.g., to the Refs.~\onlinecite{misguich02,kitaev03,vidal08} for more details on the topological aspects. 

For this work, it is sufficient to note that the unique ground state $\left|{\rm gst}\right\rangle$ of $H_{\mathbb{Z}_2}$ corresponds in the original spin model to a spin-liquid phase. The fluxes correspond to the gapped vison excitations of this spin liquid. 

\subsection{The interpolation}\label{ssec:interpol}
In this work, we follow the idea of Refs.~\onlinecite{schwandt10,poilblanc10,poilblanc11} and consider the Hamiltonian
\begin{align}
H\left( \lambda\right) = \left( 1-\lambda \right) H_{\mathbb{Z}_2} + \lambda H_{\rm QDM} ,
\label{eq:interpolation}
\end{align}
which interpolates from the exactly solvable Hamiltonian $H_{\mathbb{Z}_2}$ at $\lambda=0$ to the dimer model $H_{\rm QDM}$ at $\lambda=1$. 

If the gap of the $\mathbb{Z}_2$-phase remains open along this path, no phase transition occurs. In this case the quantum-dimer model $H_{\rm QDM}$ harbors a spin liquid as ground state. If, on the other hand, there is a phase transition signaled by a closing gap at some value $\lambda_c$ for $0<\lambda_c<1$, the investigation of the corresponding soft modes yields insights into the nature of the phase realized at $\lambda>\lambda_c$. Previous numerical results based on finite-sized system analysis \cite{schwandt10,poilblanc11} report a phase transition near the "Heisenberg point" $\lambda=1$. Our focus here is to perform a similar analysis without the constraint of finite system size.

For our analysis, we represent the Hamiltonian $H(\lambda)$ in the flux-basis Eq.~(\ref{eq:example_state}). The resulting operators involve seven flux degrees of freedom (located on one central and the six adjacent hexagons of the \kagome -lattice). Treating the flux-presence and -absence as an Ising degree of freedom we obtain the representation of the quantum dimer model in the (dual) Ising language, where the effective degrees of freedom are located on the triangular lattice formed by the hexagonal plaquettes. 

Let us conclude this section with remarks on the symmetries of the Hamiltonian $H\left( \lambda\right)$ in the flux basis. Despite the reference state breaking the translational symmetry of the lattice, flux-states respect the lattice symmetries of the underlying lattice. However, additional gauge factors enter in the corresponding transformations. E.g., for the translational symmetries within the unit cell we have
\begin{align}
T_{\vec{n}_1/2}: &\left( \begin{array}{c} P^{(1)}_{\vec{r},A}\\ P^{(1)}_{\vec{r},B} \\ P^{(1)}_{\vec{r},C} \\P^{(1)}_{\vec{r},D} \end{array}\right) \rightarrow  \left( \begin{array}{c} -{P^{(1)}_{\vec{r},D}}_{\phantom{+\vec{n}_1}}\\ \phantom{-}{P^{(1)}_{\vec{r},C}}_{\phantom{+\vec{n}_1}} \\ -P^{(1)}_{\vec{r}+\vec{n}_1,B} \\ \phantom{-}P^{(1)}_{\vec{r}+\vec{n}_1,A} \end{array}\right),\label{eq:sym_translation1}\\
T_{\vec{n}_2/2}: &\left( \begin{array}{c} P^{(1)}_{\vec{r},A}\\ P^{(1)}_{\vec{r},B} \\ P^{(1)}_{\vec{r},C} \\P^{(1)}_{\vec{r},D} \end{array}\right) \rightarrow  \left( \begin{array}{c} -{P^{(1)}_{\vec{r},C}}_{\phantom{+\vec{n}_2- \vec{n}_1}}\\ -P^{(1)}_{\vec{r}+\vec{n}_2 - \vec{n}_1,D} \\ \phantom{-}{P^{(1)}_{\vec{r}+\vec{n}_2,A}}_{\phantom{+\vec{n}_2}} \\ \phantom{-}{P^{(1)}_{\vec{r}+\vec{n}_1,B}}_{\phantom{+\vec{n}_2}} \end{array}\right),\label{eq:sym_translation2}\\
T_{\left(\vec{n}_2 - \vec{n}_1\right)/2}: &\left( \begin{array}{c} P^{(1)}_{\vec{r},A}\\ P^{(1)}_{\vec{r},B} \\ P^{(1)}_{\vec{r},C} \\P^{(1)}_{\vec{r},D} \end{array}\right) \rightarrow  \left( \begin{array}{c} -{P^{(1)}_{\vec{r},B}}_{\phantom{+\vec{n}_2-\vec{n}_1}}\\ \phantom{-}P^{(1)}_{\vec{r}+\vec{n}_2 - \vec{n}_1,A} \\ \phantom{-}P^{(1)}_{\vec{r}+\vec{n}_2 - \vec{n}_1,D} \\ -{P^{(1)}_{\vec{r},C}}_{\phantom{+\vec{n}_2-\vec{n}_1}} \end{array}\right).
\label{eq:sym_translation3}
\end{align}
These translations anti-commute with each other. This can be attributed to static gauge charges present within the unit cell as has been witnessed already in other works as, e.g., in Ref.~\onlinecite{wan13}. 

The static gauge charges have another impact: Inversion symmetry acts non-trivially on single-flux states. E.g., the action of $I_b^{\rm AC}$, the inversion about the lattice site located between the $\rm A$- and the $\rm C$-hexagon, reads:
\begin{align}
I_b^{\rm AC}: &\left( \begin{array}{c} P^{(1)}_{\vec{r},A}\\ P^{(1)}_{\vec{r},B} \\ P^{(1)}_{\vec{r},C} \\P^{(1)}_{\vec{r},D} \end{array}\right) \rightarrow  \left( \begin{array}{c} -P^{(1)}_{-\vec{r},C}\\ -P^{(1)}_{-\vec{r},D} \\ \phantom{-}P^{(1)}_{-\vec{r},A} \\ \phantom{-}P^{(1)}_{-\vec{r},B} \end{array}\right).
\label{eq:sym_inversion}
\end{align}
The non-trivial gauge part of the lattice symmetries as, e.g., $I_b^{\rm AC}$ (\ref{eq:sym_inversion}) restricts in particular the single-flux dynamics as we shall discuss in Sec.~\ref{ssec:results_1qp}. This contrasts to the less restricted interacting flux-sector as discussed in Sec.~\ref{ssec:results_2qp}.

\section{Method}\label{sec:pcut}
The low-energy spectrum of flux-Hamiltonian $H(\lambda)$ is studied by means of the perturbative continuous unitary transformations (pCUT)\cite{knetter00}. For this work, it is sufficient to consider the continuous unitary transformations as a tool, which decouples the Hilbert space into different sectors. Those sectors are investigated separately. This decoupling can be thought of as the identification of the elementary excitations of the exactly solvable Hamiltonian $H_{\mathbb{Z}_2}$ with particles, which are continuously modified to yield a quasi-particle picture. The effective model $H_{\rm eff}$ conserves the number of quasi-particles, which allows for an efficient treatment of the low particle-number sectors. So, one can formulate the continuous unitary transformation as
\begin{align}
H_{\rm eff}= U^{\phantom{\dagger}} \!\! H(\lambda)\, U^{\dagger}  = H_{\rm eff}^{\rm 0qp} + H_{\rm eff}^{\rm 1qp} + H_{\rm eff}^{\rm 2qp} + \ldots,
\label{eq:pcut_u}
\end{align}
where $H_{\rm eff}^{n {\rm qp}}$ is the effective Hamiltonian for the $n$ quasi-particle sector.

In this work, we perform this decoupling perturbatively. The interpolation parameter $\lambda$ serves here as the expansion parameter. We focus in the remainder of our work mostly on the one and two quasi-particle sectors, for which the effective Hamiltonians $H_{\rm eff}^{\rm 1qp}$ and $H_{\rm eff}^{\rm 2qp}$ are investigated respectively.
For the purpose of this work, it is sufficient to think of the pCUT as a method to perform a degenerate perturbation theory treatment of these sectors with only implicitly redefining the quasi-particle states. 
Let us note that this method can be applied to any Hamiltonian of the form $Q + \lambda \sum_{n=-N}^N T_n$, where the operator $Q$ counts the (integer) number of quasi-particles and the operators $T_n$ change the particle number by $n$.

To meet these requirements, we actually evaluate instead of $H(\lambda)$ the rescaled model $H^{\prime}(\lambda^{\prime})$, given by 
\begin{align}
H(\lambda)= & \frac{1-\lambda}{2}  \left( 2 H_{\mathbb{Z}_2} +\frac{2\lambda}{1-\lambda} H_{\rm QDM} \right) = \frac{1-\lambda}{2}H^{\prime}(\lambda^{\prime})\nonumber \\
=&\frac{1-\lambda}{2} \left( Q + \lambda^{\prime}\!\! \sum_{n=-N}^N T_n\right)\!\!\text{ with }\lambda^{\prime}=\frac{2\lambda}{1-\lambda},
\label{eq:rescaled_ham}
\end{align}
and rescale the results to be consistent with, e.g., Ref.~\onlinecite{schwandt10}.

One of the major advantages of this method is that the perturbative implementation benefits from the linked-cluster property\cite{knetter03,dusuel10} of the transformed Hamiltonian $H_{\rm eff}$. This property essentially states that only operators with connected supports contribute to the effective model $H_{\rm eff}^{n {\rm qp}}$. The size of these supports is bounded for a finite order of the perturbative treatment. Therefore the effective Hamiltonians in the thermodynamic limit can be derived on a finite-sized but large enough system. 
This fact sets this method apart from numerical methods as, e.g., exact diagonalization which are constrained by the finite sizes of study-able systems. 

Due to the linked-cluster property, we can get access to either extensive quantities as the ground-state energy or spectral differences as the energy-gaps in the thermodynamic limit. We focus mostly on the gaps in the remainder of this work.


\subsection{Free quasi-particle sector}\label{ssec:eff1qp}
The conservation of the quasi-particle number means that the effective model for the single quasi-particle sector $H_{\rm eff}^{\rm 1qp}$ is a hopping Hamiltonian. Given the effective hopping elements, this Hamiltonian can be diagonalized employing translational symmetry with the unit cell depicted in Fig.~\ref{fig:refstate}. This yields the dispersion relation $\omega(\vec{k})$. This function is thus known analytically and can therefore also be easily extrapolated beyond the perturbative parameter-range. 

\subsection{Interacting quasi-particles}\label{ssec:eff2qp}
The interaction of quasi-particles prevents a straight-forward diagonalization of the higher-number quasi-particle sectors. In this work, the two quasi-particle sector is therefore diagonalized numerically. To reduce the numerical effort, we use translational symmetry with the choice of unit-cell shown in Fig.~\ref{fig:refstate}. Thus the analysis still takes place in the thermodynamic limit as the necessary truncation for the numerical study affects only the maximal relative distance of the two quasi-particles. However, the scale of this truncation can be chosen such that the numerical results are converged.

\subsection{Observables}
Due to the implicit nature of the basis transformation, observables cannot be evaluated directly but need to be transformed by the same unitary transformation\cite{knetter03}. The resulting operators are also subjected to the linked-cluster property\cite{knetter03} and can therefore be evaluated in the same fashion as the effective Hamiltonian. We therefore have access to quantities as the ground-state expectations values and differences of expectation values of ground-state and excited states. 

\section{Results}\label{sec:results}
In this section, we will discuss our findings for the properties of the extension of the $\mathbb{Z}_2$-phase along the interpolation path (\ref{eq:interpolation}). First, we discuss the single (and hence free) quasi-particle sector in Sec.~\ref{ssec:results_1qp}. This discussion shall serve a basis, on which to discuss our results for the interacting two quasi-particle sector in Sec.~\ref{ssec:results_2qp}. We conclude this section by discussing properties of the relevant low-energy states in Sec.~\ref{ssec:obs}.

\subsection{Free quasi-particle sector}\label{ssec:results_1qp}
\begin{figure}[ht]%
\includegraphics[width=.95\columnwidth]{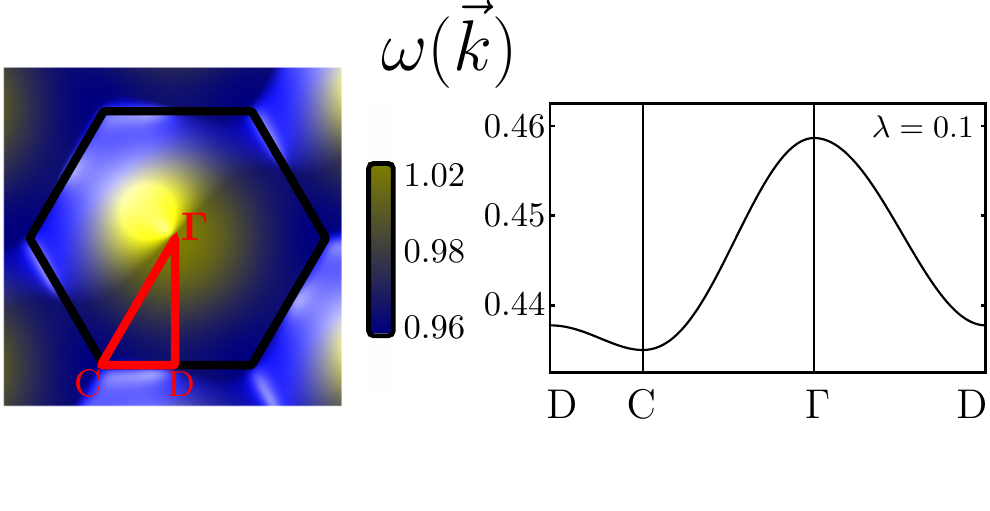}%
\vspace{-0.75cm}
\caption{On the left side, we show the dispersion $\omega(\vec{k})$. The corresponding hopping terms have been determined up to order four. We find extrema of the dispersion at the points $\Gamma$, $\rm C$, and $\rm D$ (we chose these labels consistent with Refs.~\onlinecite{poilblanc10,poilblanc11}). 
\\
On the right side, we also show a cut through the $\vec{k}$-space along the path $D-C-\Gamma-D$. We find the absolute minimum of $\omega(\vec{k})$ at point $\rm C$. }%
\label{fig:dispersion_kcut}%
\end{figure}

We determine the effective single-flux Hamiltonian $H_{\rm eff}^{\rm 1qp}$ up to order four (The matrix elements are given in App.~\ref{app:hopping1qp}). Due to inversion symmetry properties Eq.~(\ref{eq:sym_inversion}), matrix elements connecting single-flux states on different sub-lattices (the different hexagons labeled by $\rm A$ - $\rm D$ in Fig.~\ref{fig:refstate}) vanish. So we are left with one model of hopping particles for each sub-lattice respectively. These different models are related by the inner unit cell translations discussed in Eqs.~(\ref{eq:sym_translation1}-\ref{eq:sym_translation3}). Therefore, we find four degenerate single-flux bands distinguished by their respective sub-lattice index. We show our results for the dispersion $\omega(\vec{k})$ in Fig.~\ref{fig:dispersion_kcut}.

Because of the relatively low orders in our series expansion, we discuss our results first in the perturbative regime ($\lambda = 0.1$), characterized by a small bandwidth (compared to the spectral gap). We find the minima of the dispersion $\omega(\vec{k})$ at the corners of the Brillouin zone at $\vec{k}_{\pm}=\pm(\frac{2\pi}{3},-\frac{2\pi}{3})$, i.e. the $\vec{k}$-points corresponding to the points $\rm B,C$ in Ref.~\onlinecite{schwandt10}. 

\begin{figure}[ht]%
\includegraphics[width=.75\columnwidth]{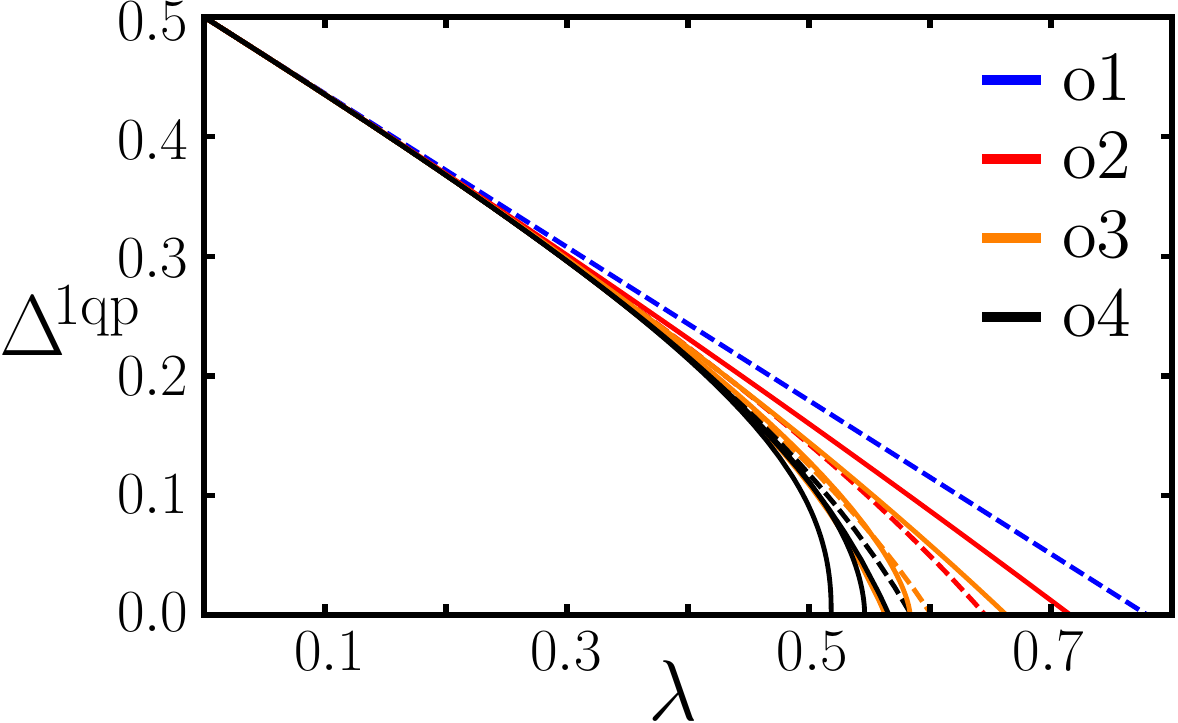}%
\caption{The free quasi-particle gap $\Delta^{\rm 1qp}(\lambda)$ (dashed) as well as different extrapolants (solid) are plotted from order one (blue) to order four (black). The closure of the gap shows the tendency to occur for smaller values of $\lambda$ with increasing order (cf. App.~\ref{app:extrapolants_disp} for details). Due to the low orders used, the results are not converged; however the highest-order results are consist with a value for $\lambda_c = 0.57$.}%
\label{fig:gap1qp}%
\end{figure}

To obtain an estimate of the extend of the $\mathbb{Z}_2$-phase, we determine the location $\lambda_c$, where the gap {$\Delta^{1{\rm qp}}=\omega(\vec{k}_{\pm})$} closes. The gap as well as some extrapolations are shown in Fig.~\ref{fig:gap1qp} for different orders. For increasing order, the closure occurs at smaller values of $\lambda$. The orders obtained in this work are not high enough to yield converged results without further assumptions, however the highest-order results are consistent with $\lambda_c = 0.57$. 

This estimate is an upper bound for the location of the phase transition. So, we can already conclude that the numerical value of $\lambda_c = 0.94$ reported in Ref.~\onlinecite{schwandt10} is not consistent with our findings. We defer any further comparisons after we discussed the interactions in the following section.

\subsection{Interacting quasi-particles}\label{ssec:results_2qp}
We estimate the role of the interactions by analyzing the interacting two quasi-particle sector. We determine the effective Hamiltonian in the sector of two quasi-particles $H_{\rm eff}^{\rm 2qp}$ up to order three. For the further analysis, we make use of the translational symmetry and study the sectors for different values of the total momentum $\vec{k}$ separately. We consider in the remaining parts of this work two quasi-particle states with the relative distance of at most $d_{\rm max}=50$ (in units of $\left|\vec{\boldsymbol{n}}_1\right|$), for which we find converged results. 

\begin{figure}[ht]%
\includegraphics[width=.75\columnwidth]{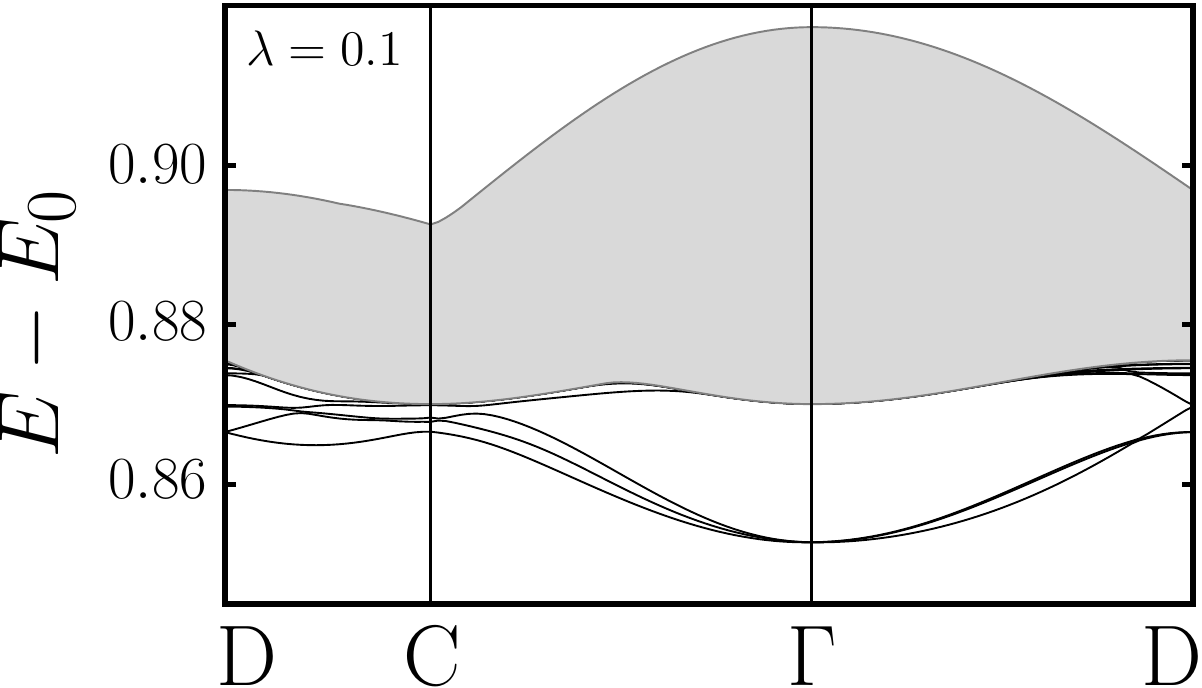}%
\caption{The energy spectrum of the interacting two quasi-particle sector is shown along the path $D-C-\Gamma-D$ in the Brillouin zone defined in Fig.~\ref{fig:dispersion_kcut}. The gray-shaded area depicts the energy window available for the two quasi-particle continuum. Below the continuum, we show the dispersion of bound states. We find the absolute minimum of the energy here at the $\Gamma$-point, where the lowest level is three-fold degenerate.}%
\label{fig:spectrum_2qp}%
\end{figure}

Similar to the previous section, we discuss first the spectrum in the perturbative regime shown in Fig.~\ref{fig:spectrum_2qp}. 
Here, we depict the energy spectrum along the path in the Brillouin-zone shown in Fig.~\ref{fig:dispersion_kcut}. The gray-shaded area denotes the energy-window covered by the continuum of two non-interacting quasi-particles obtained from the free quasi-particle dispersion $\omega(\vec{k})$ at the same order. Additionally to these continuum states, we display bound states below the continuum, forming a new low-energy manifold of states. The states lowest in energy are the three-fold degenerate states at $\vec{k}_0=(0,0)$. 

\begin{figure}[ht]%
\includegraphics[width=.75\columnwidth]{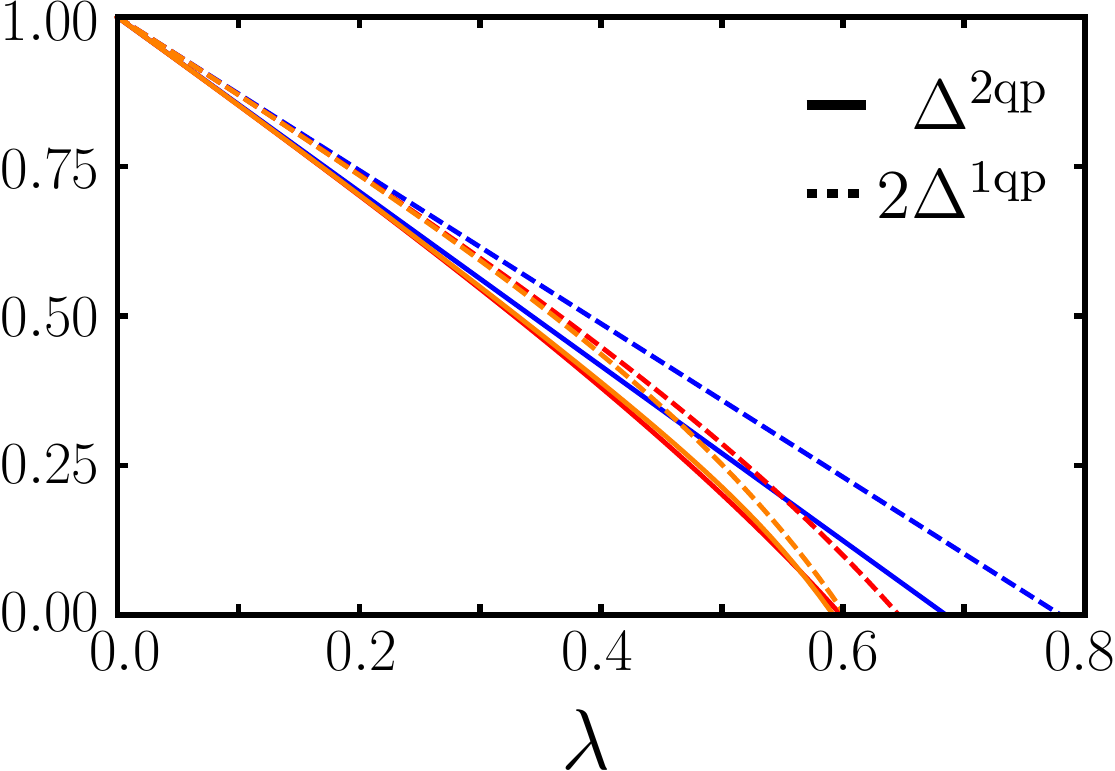}%
\caption{The lowest two-particle energy (solid) is shown as well as the energy of the lower continuum-edge (dashed) for the orders one (blue), two (red), and three (orange). For each order there is a finite energy gap between these two values for all $\lambda>0$. So we conclude that the bound states exist for all of these values. As their energy is similar to the one of the continuum edge, we estimate $\lambda_c$ by the better controlled single-particle properties. }%
\label{fig:gap2qp}%
\end{figure}

Their energy represents the two quasi-particle gap $\Delta^{\rm 2qp}(\lambda)$, which is shown in Fig.~\ref{fig:gap2qp} together with the energy of the lower edge of the continuum at the same order. 
The bound states exist for any finite value of $\lambda$ and therefore their energy is always lower than the continuum. We assume that the effective models $H_{\rm eff}^{\rm 1qp}$, $H_{\rm eff}^{\rm 2qp}$ capture the relevant low-energy physics (the ranges of the terms in the effective models cover twice the range associated to the unit-cells discussed in Sec.~\ref{ssec:obs}) and therefore the trend seen in the higher-order free quasi-particle results holds also for the interacting particles. 
However, the results are not fully converged up to maximal order obtained here. 
Given that the difference between the lower continuum edge and bound state gap is small, we use the lower continuum edge $2\Delta^{\rm 1qp}$, which is analytically available at higher orders, to estimate the location of the phase transition out of the $\mathbb{Z}_2$ spin-liquid to be $\lambda_c=0.57$. 

In order to investigate the order arising beyond this transition at $\lambda_c$, let us discuss the structure of the low-energy states.

\subsection{Properties of the low-energy states}\label{ssec:obs}
To obtain insights into the order arising at the transition point, where the gap of the excitations of the $\mathbb{Z}_2$-phase closes, we investigate the properties of the low-energy states introduced in Sec.~\ref{ssec:results_2qp}. In parallel, we discuss the characteristics resulting from the free-particle physics reflected in the physics of the two quasi-particle continuum. 

\begin{figure}[ht]%
\includegraphics[width=.5\columnwidth]{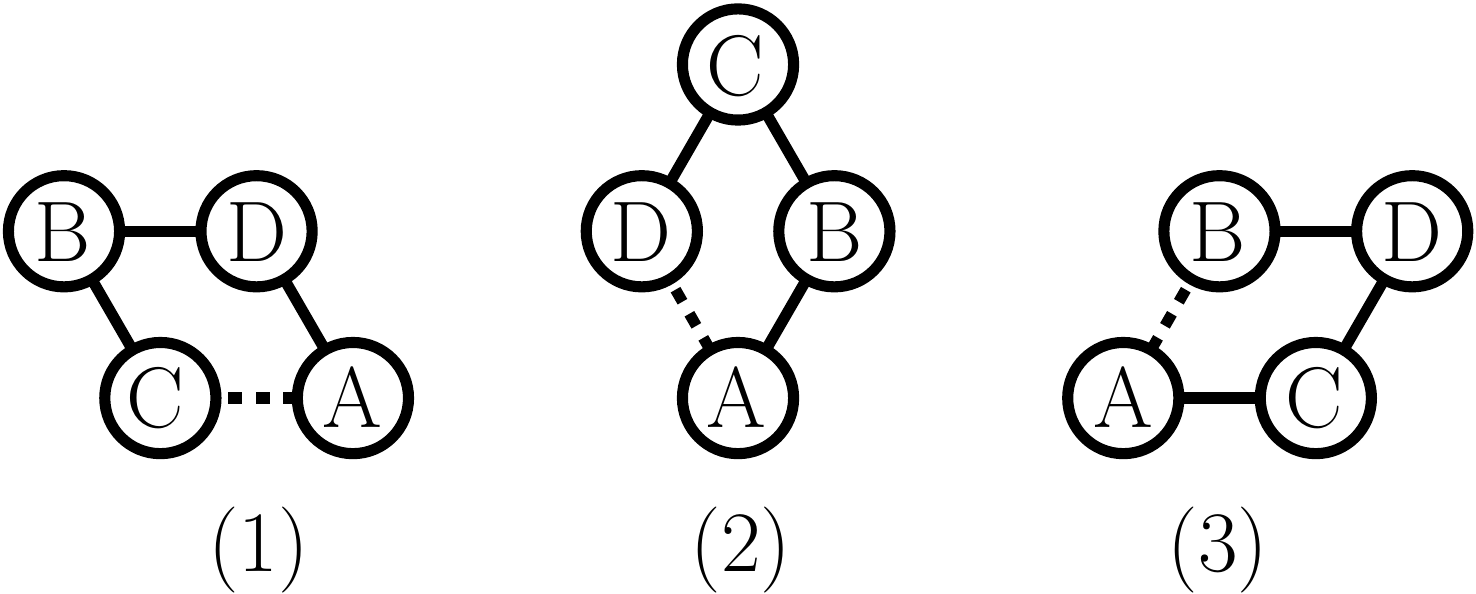}%
\caption{Representation of the nearest-neighbor part of the three bound states. Fig.~(i) depicts the state $\left|{\rm bst}_i\right\rangle$. Letters indicate the sub-lattice index. Two indices connected by a solid line indicate a nearest-neighbor pair of quasi-fluxes on plaquettes indicated by the respective sub-lattice indices. A dashed line indicates a state with an amplitude of opposite sign.}%
\label{fig:bst_nnsketch}%
\end{figure}

The lowest energy in the two quasi-particle sector is found at the $\Gamma$-point ($\vec{k}=(0,0)$). There are three degenerate states, labeled by $\left| {\rm bst}_i\right\rangle$, $i=1,2,3$ in the following. In Fig.~\ref{fig:bst_nnsketch}, we show a projection onto the nearest-neighbor quasi-flux pairs of the these three orthogonal states.

Let us mention here that the states $\left| {\rm bst}_i\right\rangle$ are eigenstates of the inner unit cell translations (\ref{eq:sym_translation1}-\ref{eq:sym_translation3}) and inversions (\ref{eq:sym_inversion}) with eigenvalues $\pm 1$ and map onto each other under the action of the rotational symmetry. We list the detailed action of the other lattice-symmetries in App.~\ref{app:symmetry_actions}.

Assuming that these states are the relevant low-energy states at the phase transition, we can reason along the following line of thought:
There are no other states degenerate with the states $\left| {\rm bst}_i\right\rangle$ for a different $\vec{k}$-value. Therefore a spontaneous breaking of the translational symmetry and the formation of larger unit cells cannot occur by condensation of these states. 

\begin{figure}[ht]%
\begin{minipage}{\columnwidth}
\begin{minipage}{.05\columnwidth}
\quad
\end{minipage}
\begin{minipage}{.4\columnwidth}
\includegraphics[width=.75\columnwidth]{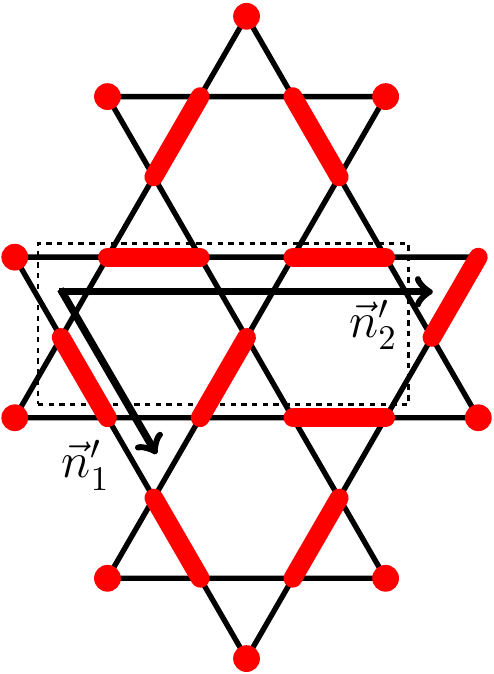}%
\end{minipage}
\begin{minipage}{.05\columnwidth}
\quad
\end{minipage}
\begin{minipage}{.4\columnwidth}
\includegraphics[width=.95\columnwidth]{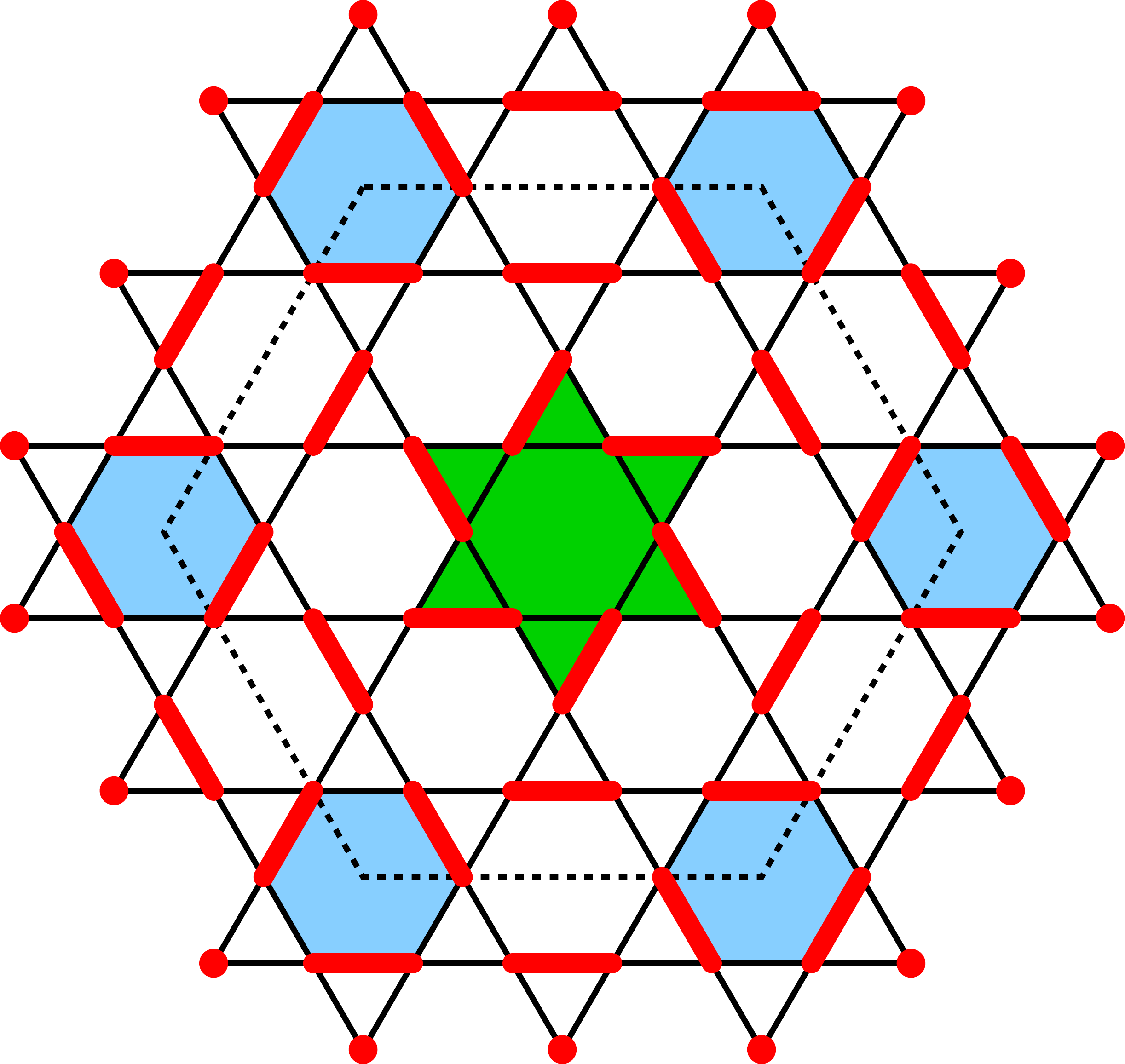}%
\end{minipage}
\begin{minipage}{.05\columnwidth}
\quad
\end{minipage}
\end{minipage}
\caption{On the left, we depict a typical dimer configuration being present in the local wave packet $\left|\psi_3\right\rangle$ (\ref{eq:wavepakets}) at $\lambda\approx 0$. To represent this wave packet, we truncate the maximal relative distance of the quasi-particle states to $1$ (in units of $\left|\vec{n}_1\right|$). The corresponding bound state $\left|{\rm bst}_3\right\rangle$ can be shown to have the 6-site unit cell depicted by the dashed lines.
\\
On the right, we show a typical dimer configuration for the 36-sites unit cell of the valence bond solid arising in the free quasi-particle condensation. The hexagons (blue) and the pinning-wheel structure (green) are resonating (cf. Refs.~\onlinecite{schwandt10,poilblanc10} for a more detailed discussion).}%
\label{fig:dimer_pic}%
\end{figure}

However, each state $\left|{\rm bst}_i\right\rangle$ is the eigenstate of one inner unit cell translations (\ref{eq:sym_translation1}-\ref{eq:sym_translation3}) with eigenvalue $+1$, respectively. This allows us to split for each bound state the twelve site unit cell shown in Fig.~\ref{fig:refstate} into a six site unit cell as, e.g., depicted in Fig.~\ref{fig:dimer_pic}.

Thus we can recast the bound states as 
\begin{align}
\left| {\rm bst}_i\right\rangle = \mathcal{N}_i \sum_{\vec{n}^{\prime}} \left|\psi_i\right\rangle_{\vec{n}^{\prime}},
\label{eq:wavepakets}
\end{align}
where the $\left|\psi_i\right\rangle_{\vec{n}^{\prime}}$ are local wave packets centered around the six site unit cell with coordinates $\vec{n}^{\prime}$. 

At $\lambda\approx 0$, where particles and quasi-particle picture are essentially identical, we can use Eq.~(\ref{eq:example_state}) to obtain typical dimer patterns present in the bound state as depicted in Fig.~\ref{fig:dimer_pic}. 
These states are adiabatically connected to the condensing patterns at the phase transition. However, due to the fact that two quasi-particle states with increasing distance contribute more for increasing values of $\lambda$, we expect longer-ranged correlations to dampen expectation values of any short-ranged dimer configuration. This dampening renders the identification of the resulting phase from such typical patterns more subtle.

In contrast to the bound-state condensation described above, the condensation of free quasi-particles allows for spontaneous breaking of the translational symmetry:
being formed by independent one quasi-particle states with momentum $\pm(\frac{2\pi}{3},-\frac{2\pi}{3})$, the lower continuum edge at the $\vec{k}$-points $\vec{k}_{\pm}=\pm(\frac{2\pi}{3},-\frac{2\pi}{3})$ and $\vec{k}_0=(0,0)$ is degenerate and allows for the formation a valence bond solid having a thirty-six site unit cell\cite{marston91,nikolic03,singh07,schwandt10,poilblanc10,poilblanc11}. Using again Eq.~(\ref{eq:example_state}), we depict a typical state for this case in Fig.~\ref{fig:dimer_pic}.

However, as for $\lambda\neq 0$, the quasi-particle states differ from the flux eigenstates. Thus, we cannot determine the dimer configurations via Eq.~(\ref{eq:example_state}) for larger values of $\lambda$. 
To capture nevertheless the short-ranged correlations between different dimers, we transform the kinetic and potential terms in the quantum-dimer Hamiltonian defined in Eq.~(\ref{eq:ham_qdm}). 
More specifically, we determine the difference between the expectation values of the ground state and the excited states.
These quantities are spectral differences and are therefore accessible as discussed in Sec.~\ref{sec:pcut}. We determine the transformed observables up to order one and determine the (degenerate) expectation values for the states $\left|{\rm bst}_i\right\rangle$.
Due to the low order of the observable-expansion, we restrict our discussion here to the perturbative regime $\lambda\ll\lambda_c$. However, the trends seen here shall yield some insight into the adiabatic development of dimer correlations for increasing $\lambda$. 

\begin{figure}[ht]%
\includegraphics[width=\columnwidth]{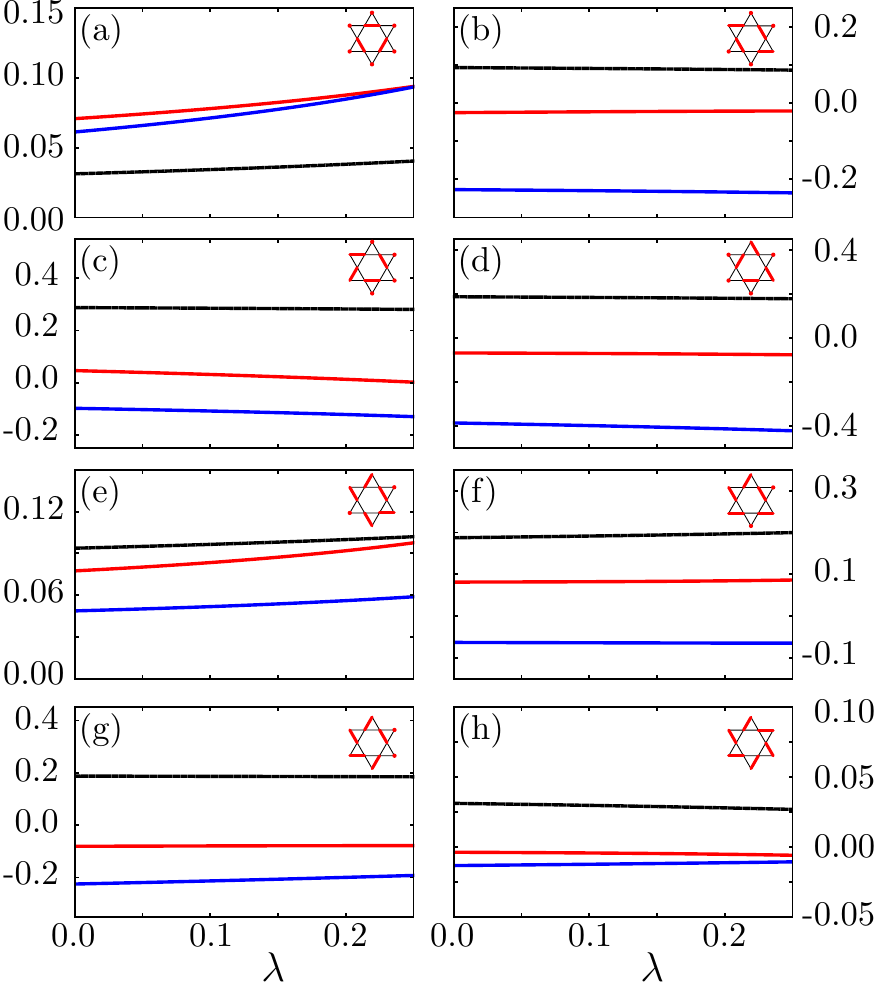}%
\caption{The expectation value for the kinetic (blue) and potential (red) terms (see text) in dependence of $\lambda$ for the dimer configuration depicted within each plot. The black line denotes the ground state expectation value per unit cell (being the same for kinetic and potential terms) for comparison. We restrict ourselves to the perturbative regime as we do not extrapolate the data. }%
\label{fig:obs_expectations_A}%
\end{figure}

In Fig.~\ref{fig:obs_expectations_A}, we show the ground-state expectation values for of the quasi-flux free ground state (per unit cell). The expectation values for the kinetic and potential terms coincide, as the ground state remains a superposition of perfectly resonating dimer configurations at this order. The situation is different for the excited states. We find two different families of observables: For the 3-dimer (\ref{fig:obs_expectations_A}$({\rm a})$) and 6-dimer (\ref{fig:obs_expectations_A}$({\rm h})$) observables, the expectation values of kinetic (blue) and potential (red) terms are close to each other, indicating a resonating contribution of hexagons and pinning-wheels. We attribute this to the long-distance tails of the bound states, which behave essentially as free quasi-particles. Those states host these resonating plaquettes as can be seen from Fig.~\ref{fig:dimer_pic}. We expect those resonating contributions to be present in the condensed state, however their magnitude remains small compared to those of the other family of observables.

The other terms, shown in Fig.~\ref{fig:obs_expectations_A}$({\rm b})$-Fig.~\ref{fig:obs_expectations_A}$({\rm g})$, involve 4-dimer and 5-dimer states. We see for those sizable differences between kinetic and potential terms. These differences correspond to non-resonating contributions forming crystalline patterns in terms of dimers. The main contributions here stem from the short-distance interacting quasi-particle states. The most notable difference is found for the diamond \ref{fig:obs_expectations_A}$({\rm b})$ and trapezoidal \ref{fig:obs_expectations_A}$({\rm d})$ -shaped configurations, indicating that these patterns shall be most prominent in a solid arising beyond the phase transition.

We see that this analysis cannot give us quantitative results due to the low order of the expansion, but still yields insights into the structure of the condensed state.

\section{Discussion and conclusion}\label{sec:conclusion}
In this work, we investigate the low-energy states relevant to an interpolation between a gapped $\mathbb{Z}_2$ spin liquid and an effective model describing the low-energy physics of the Heisenberg anti-ferromagnet on the \kagome -lattice. 
A perturbative approach is employed to decouple different sectors of the Hilbert space, namely the free quasi-particle sector and the interacting two quasi-particle sector. These sectors are investigated separately. This approach enables us to obtain results directly in the thermodynamic limit and at the same time treat the quasi-particle interactions in a controlled fashion.

Studying the low-energy physics of the interacting two quasi-particle sector, we find the phase transition to be driven by the condensation of quasi-flux bound states. From the analysis of the low-energy dispersion of those bound states, we conclude that the phase resulting from the such a condensation has a six site unit cell as shown in Fig.~\ref{fig:dimer_pic}. In particular we can exclude a unit cell with larger size. The states found in our analysis break spontaneously the six-fold rotational symmetry of the underlying lattice. 
The location of the phase transition out of the topological phase is estimated to occur at $\lambda_c\approx 0.57$, whereas previous numerical studies\cite{schwandt10,poilblanc10} found it much closer to the Heisenberg point $\lambda=1$ (within the same parametrization). 

Interestingly, we find the properties of the free quasi-particle condensation to agree with the numerical analysis\cite{schwandt10,poilblanc10}. We attribute the differences in our analysis to the finite-size limitations of the numerical studies, as the bound states discussed in Sec.~\ref{ssec:results_2qp} form on length scales not accessible in those.

Our different findings in the free and the interacting quasi-particle sector therefore provide an example that the explicit study of interacting quasi-particles (not only within perturbative approaches) may yield additional insights not accessible otherwise. 

Performing the perturbative transformation also for different observables, we are able to obtain expectation values of the condensing bound states to yield further insight into their structure. The most dominant contribution to crystalline patterns within these bound states stems from diamond- and trapezoidal- shaped dimer configurations as they show the least resonant behavior. 
Therefore, we expect a valence bond solid state with those patterns to arise after the condensation transition. \\
Recently, alternative derivations of the quantum-dimer model including longer-ranged singlets show that the corresponding kinetic and potential terms can even be further favored\cite{rousochatzakis14}. This may provide the possibility that those states are even stabilized more clearly in the original Heisenberg model.

With our work, we add more insights into the phase arising in the low-energy description of the anti-ferromagnetic Heisenberg model on the \kagome -lattice. Our analysis suggests that a phase transition terminates the $\mathbb{Z}_2$ topological phase before reaching the Heisenberg point. We find this phase transition is driven by the condensation of interacting quasi-fluxes.

Understanding the behavior of the topological excitations allows us to further shed light onto the low-energy properties of the condensed phase, which is found to break the rotational symmetry spontaneously and has a six site unit cell.

\acknowledgments
We would like to thank J.~Vidal, D.~Poilblanc, G.~Misguich, Y.~Wan, and I.~Rousochatzakis for fruitful discussions. This research has in parts been carried out while enjoying the hospitality of the Perimeter Institute.

\appendix
\section{Results for the free quasi-particle sector}\label{app:hopping1qp}
We determine the effective hopping elements $t_{\vec{n}}$, from which one can obtain the dispersion $\omega(\vec{k})$ as 
\begin{align}
\omega(\vec{k})=\sum_{\vec{n}}e^{\mathrm{i}\vec{k}\vec{n}}t_{\vec{n}}.
\label{eq:}
\end{align}
The vector $\vec{n}$ describes the position of the unit cells and we drop here the sub-lattice index as the hopping terms are diagonal and degenerate in this variable.
We determine the hopping elements in terms of the rescaled interpolation parameter $\lambda^{\prime}$ defined in Eq.~(\ref{eq:rescaled_ham}).

The non-zero hopping elements read up to order four:
\begin{align}
t_{(0,0)}=&\ \ 1-\frac{7384}{1785}{\lambda^{\prime}}^{\phantom{1}}-\frac{739867952}{9558675}{\lambda^{\prime}}^2\nonumber\\&
-\frac{365891064704}{210644875}{\lambda^{\prime}}^3\nonumber\\&
-\frac{151716416343032361568}{4111572049003125}{\lambda^{\prime}}^4,\\
t_{(0,1)}=&	\frac{972}{595}{\lambda^{\prime}}^{\phantom{1}}+\frac{809136}{50575} {\lambda^{\prime}}^2+\frac{12018333392}{812487375}{\lambda^{\prime}}^3\nonumber\\&
-\frac{152780635630652096}{30456089251875}{\lambda^{\prime}}^4,\\
t_{(0,2)}=&	-\frac{944784}{354025}{\lambda^{\prime}}^2-\frac{3462423408}{210644875}{\lambda^{\prime}}^3\nonumber\\&
+\frac{27536006198016}{125333700625} {\lambda^{\prime}}^4,\\
t_{(0,3)}=&	\frac{1836660096}{210644875} {\lambda^{\prime}}^3+\frac{39182082048}{
   5013348025} {\lambda^{\prime}}^4,\\
t_{(0,4)}=&-\frac{892616806656}{25066740125} {\lambda^{\prime}}^4,\\
t_{(1,1)}=&-\frac{1889568}{354025}{\lambda^{\prime}}^2-\frac{943720416}{12390875}{\lambda^{\prime}}^3\nonumber\\&
-\frac{4867412039808}{7372570625}{\lambda^{\prime}}^4,\\
t_{(1,2)}=&\frac{5509980288}{210644875} {\lambda^{\prime}}^3+ \frac{52585417728}{73080875}{\lambda^{\prime}}^4,\\
t_{(1,3)}=&-\frac{3570467226624}{25066740125} {\lambda^{\prime}}^4,\\
t_{(2,2)}=&-\frac{5355700839936}{25066740125} {\lambda^{\prime}}^4.
\label{eq:hopping_amplitudes}
\end{align}

\section{Extrapolant data for the one quasi-particle gap}\label{app:extrapolants_disp}
In Tab.~\ref{tab:extrapols}, we give the data obtained from the extrapolation of the one quasi-particle gap $\Delta^{1\rm{qp}}$ shown in Fig.~\ref{fig:gap1qp}: In order to determine the location of the gap closure $\lambda_c$, we use the bare series as well as Pad\'e- and dlog-Pad\'e-extrapolations. The Pad\'e-extrapolation assumes analytic behavior in the parameter $\lambda$, whereas the dlog-Pad\'e-extrapolation assumes algebraic behavior near the closing of the gap $\Delta^{1\rm{qp}}$. 

\begin{table}[hp]%
\begin{tabular}{l | c | c | c |}
order & \quad funct. \quad & \quad \quad $\lambda_c$ \quad \quad & \quad exp. \quad\\
\hline
order 1		& \ \ $\rm{bare}$ & $0.78$ & - \\
\hline
order 2		& \ \ $\rm{bare}$ & $0.72$ & - \\
& $\phantom{\rm{d}\!}\left[ 1,1\right]$ & $0.72$ & - \\
\hline
order 3		& \ \ $\rm{bare}$ & $0.68$ & - \\
& $\rm{d}\!\left[ 1,1\right]$ & $0.58$ & $0.66$ \\
& $\phantom{\rm{d}\!}\left[ 1,2\right]$ & $0.66$ & - \\
& $\phantom{\rm{d}\!}\left[ 2,1\right]$ & $0.56$ & - \\
\hline
order 4		& \ \ $\rm{bare}$ & $0.64$ & - \\
& $\rm{d}\!\left[ 2,1\right]$ & $0.55$ & $0.54$ \\
& $\rm{d}\!\left[ 1,2\right]$ & $0.52$ & $0.42$ \\
& $\phantom{\rm{d}\!}\left[ 1,3\right]$ & $0.63$ & - \\
& $\phantom{\rm{d}\!}\left[ 2,2\right]$ & $0.68$ & - \\
& $\phantom{\rm{d}\!}\left[ 3,1\right]$ & $0.57$ & - \\
\hline
\end{tabular}
\caption{Here we give the results $\lambda_c$ for the closing of the gap $\Delta^{\rm 1qp}$ shown in Fig.~\ref{fig:gap1qp}. We show the results for the bare series as well as for the Pad\'e-extrapolations $\left[m,n\right]$ and the dlog-Pad\'e-extrapolations $\rm{d}\!\left[ m,n\right]$. For the latter, we also give for completeness the exponent associated to the algebraic gap-closing behavior. However, due to the low orders considered, these values are not expected to be converged.}
\label{tab:extrapols}
\end{table}

The results show a trend towards smaller values of $\lambda_c$ as the order increases. Due to the low orders available in our expansion, the estimates from different extrapolations allow for an relatively large window of possible values of $\lambda_c$. At the highest order the results are found in the window of $0.52\leq \lambda_c \leq 0.68$. Here, we report the central value of the extrapolations of $\lambda_c\approx 0.57$.

\section{Symmetry Actions}\label{app:symmetry_actions}
In Tab.~\ref{tab:bst_sym}, we list the transformation of the bound state manifold under the action of the lattice-symmetries.

Additionally to the inner unit cell translations (\ref{eq:sym_translation1}-\ref{eq:sym_translation3}) and inversion (\ref{eq:sym_inversion}), we have the inversion about the center of a hexagon $I_h$,
the six-fold rotation $R_6$ abound the center of a hexagonal plaquette (we choose here plaquette $A$ in the unit cell depicted in Fig.~\ref{fig:refstate}) and the three-fold rotation $R_3$ about the center of a triangle. Other lattice-symmetries are the reflection about the $\vec{n}_2$-axis $M_{\vec{n}_2}^{\rm AC}$ (we choose here the axis positioned on the line of ${\rm A}- {\rm C}$ plaquettes) and the reflection about the $(\vec{n}_1+\vec{n}_2)$-axis $M_{\vec{n}_1+\vec{n}_2}^{\rm AB}$ (we choose here the axis positioned on the line of ${\rm A}- {\rm B}$ plaquettes). The other symmetry actions can be obtained by combinations of the mentioned ones.
\begin{table}[hp]%
\begin{tabular}{l|ccc|}
 &\quad$\phantom{-}\left|{\rm bst}_1\right\rangle$ & \quad$\phantom{-}\left|{\rm bst}_2\right\rangle$ & \quad$\phantom{-}\left|{\rm bst}_3\right\rangle$\\ \hline
$T_{\vec{n}_1/2}$& \quad$-\left|{\rm bst}_1\right\rangle$ & \quad$-\left|{\rm bst}_2\right\rangle$ & \quad$\phantom{-}\left|{\rm bst}_3\right\rangle$ \\
$T_{\vec{n}_2/2}$& \quad$-\left|{\rm bst}_1\right\rangle$ & \quad$\phantom{-}\left|{\rm bst}_2\right\rangle$ & \quad$-\left|{\rm bst}_3\right\rangle$ \\
$T_{(\vec{n}_2-\vec{n}_1)/2}$& \quad$\phantom{-}\left|{\rm bst}_1\right\rangle$ & \quad$-\left|{\rm bst}_2\right\rangle$ & \quad$-\left|{\rm bst}_3\right\rangle$ \\
$I_h$& \quad$\phantom{-}\left|{\rm bst}_1\right\rangle$ & \quad$\phantom{-}\left|{\rm bst}_2\right\rangle$ & \quad$\phantom{-}\left|{\rm bst}_3\right\rangle$ \\
$I_s^{\rm AC}$& \quad$-\left|{\rm bst}_1\right\rangle$ & \quad$\phantom{-}\left|{\rm bst}_2\right\rangle$ & \quad$-\left|{\rm bst}_3\right\rangle$\\
$R_6$& \quad$\phantom{-}\left|{\rm bst}_3\right\rangle$ & \quad$\phantom{-}\left|{\rm bst}_1\right\rangle$ & \quad$\phantom{-}\left|{\rm bst}_2\right\rangle$ \\
$R_3$& \quad$\phantom{-}\left|{\rm bst}_2\right\rangle$ & \quad$\phantom{-}\left|{\rm bst}_3\right\rangle$ & \quad$\phantom{-}\left|{\rm bst}_1\right\rangle$ \\
$M_{\vec{n}_2}^{\rm AC}$& \quad$\phantom{-}\left|{\rm bst}_3\right\rangle$ & \quad$\phantom{-}\left|{\rm bst}_2\right\rangle$ & \quad$\phantom{-}\left|{\rm bst}_1\right\rangle$ \\
$M_{(\vec{n}_1+\vec{n}_2)/2}^{\rm AB}$& \quad$\phantom{-}\left|{\rm bst}_1\right\rangle$ & \quad$\phantom{-}\left|{\rm bst}_3\right\rangle$ & \quad$\phantom{-}\left|{\rm bst}_2\right\rangle$ \\
\hline
\end{tabular}
\caption{Action of the symmetry operators on the bound states depicted in Fig.~\ref{fig:bst_nnsketch}. These statements are independent of the actual value of $\lambda$.}
\label{tab:bst_sym}
\end{table}

\end{document}